\documentclass[aps,prd,reprint]{revtex4-2}
\usepackage{graphicx}   % for figures

\begin{document}

\title{Electromagnetic momentum in the Aharonov-Bohm quantum interference experiment from a physical perspective} 
\author{Ashok K. Singal}
\email{ashokkumar.singal@gmail.com}
\affiliation{Astronomy and Astrophysics Division, Physical Research Laboratory,
Navrangpura, Ahmedabad - 380 009, India }
\date{\today}
\begin{abstract}
In the Aharonov-Bohm setup, a double-slit experiment, when a long but thin solenoid of current is introduced between the two coherent beams of electrons behind the slits, an extra phase difference between the interfering beams appears, as shown by a shift in the interference pattern. 
This mysterious effect, purportedly arises owing to an electromagnetic momentum, attributed to the presence of a vector potential at the location of either beam, due to the solenoid of current even when the magnetic field is zero outside the solenoid. It has remained a puzzle, how mere potential, thought to be just a mathematical tool for calculating electromagnetic field, can give rise to  electromagnetic momentum in a system. Experimentally the  effect has been amply verified, with hardly any doubts that the observed effect is real.
A satisfactory physical explanation of the existence of momentum, at least under the aegis of classical electromagnetism, is still missing since inception of the idea more than half a century back. We show here the presence of electromagnetic momentum in the product of the drift velocities of the current-carrying charges within the solenoid and the mass equivalent of their potential energies in the electric field of the external charges.
\end{abstract}
\maketitle
%%%%%%%%%%%%%%%%%%%%%%%%%%%%%%%%%%%%%%%%%%%%%%%%%
\section{Introduction}
In a double-slit interference experiment of quantum mechanics, a pair of coherent beams of electrons passing through the two slits form an interference pattern on the observing screen \cite{Po61,Fe65,Wi71,Me98}. However, when a long but thin solenoid of electric current is introduced behind the slits between the two electron beams, a shift in the interference pattern on the screen is observed, which is known as Aharonov-Bohm (AB) effect \cite{Ah59}. This effect is sometimes called Ehrenberg-Siday-Aharonov-Bohm effect as the idea was mooted apparently a decade earlier \cite{Er49}. Curiously, the shift in the interference pattern takes place due to the influence of the solenoid of current even when there exists no magnetic (or electric) field at the location of the two interfering beams, outside the solenoid.

The AB effect is a quantum physics phenomenon where a phase shift occurs in the wave function of a charged particle in the presence of an electromagnetic (EM) vector potential even when the  EM fields are absent at the location of the charge, This demonstrates apparently the physical reality of the vector potential in quantum mechanics, contrary to the view in classical electrodynamics that potentials are merely mathematical tools without a physical significance. Moreover their values could be arbitrary to some extent, depending on the choice of gauge. The AB effect supposedly arises because a charged particle in the presence of a vector potential is a source of an EM momentum, responsible for the wave function of the charged particle to acquire an extra phase, which. in turn, is illustrated by a shift in the interference pattern in the two-slit experiment. 

An extra phase difference between the two interfering beams purportedly arises owing to an EM momentum, attributed to the presence of a vector potential at the location of either beam, due to the solenoid of current even when there exists no magnetic field outside the solenoid. The choice of gauge for the vector potential, as will be discussed later, however, does not affect the extra phase difference between the interfering beams or the shift in their interference pattern on the screen. 
However, it is not obvious where the EM momentum is present in the system, especially when no electric or magnetic fields are present at the locations of external charges. 
%It is generally argued that the AB effect illustrates the physicality of EM potentials, in quantum mechanics. while classically it was thought that only the EM fields have physical reality, while the EM potentials are purely mathematical tools to calculate fields.

The  first experimental confirmation of the AB effect came soon after the prediction \cite{Ch60} and it has since been amply verified using clever experimental setups \cite{To86,Os86}, leaving hardly any doubts that the observed effect is real.
However, on the theoretical side the picture is not so clear and a satisfactory physical explanation of the existence of momentum, at least under the aegis of classical electromagnetism, is still missing since inception of the idea more than sixty years back. 
It has remained an enigma, how just potential, thought to be a mere mathematical tool for calculating EM field, can give rise to an EM momentum in a system, in the absence of the field itself. 

We show here how a small current loop, part of the solenoid considered to be a stack of many such current loops, which is a source of a finite vector potential at the location of an external electric charge, gives rise to an EM momentum. In order to resolve the enigma of the elusive EM momentum, one usually focuses attention on the external electron and attempts to study the influence of the vector potential due to the solenoid at it. However, so far this has not yielded successful results. As we shall show one should instead be looking at the effect of the external charge on the current-carrying charges within the solenoid. We shall that way demonstrate that the momentum is manifested in the product of the drift velocities of the current-carrying charges and the mass equivalent of their potential energies in the electric field of the external charge. A pair of equal charges lying at symmetrical opposite locations outside the current loop gives rise to an equal and opposite momentum in the system. This elusive, equal and opposite momentum for a pair of charges is reflected through an extra phase difference between the interfering electron beams in the double-slit experiment, when a long but thin solenoid of current, equivalent to a stack of large numbers of current loops, is introduced between the electron beams. 
 The choice of gauge for the vector potential, as will be discussed later, however, does not affect the extra phase difference between the interfering beams or the shift in their interference pattern.  

A viable explanation of this curious phenomenon in terms of such a subtle momentum under the aegis of classical electromagnetism, has been coveted since inception of the idea more than 60 years back. 
The presence of an EM momentum to have a classical origin in the system has been attempted for a {\em moving} charge \cite{Bo08}. However, as we shall see, 
% according to Eq.~({\ref{eq:84a.3}) 
one should be able to account for the EM momentum in the system {\em even for a stationary charge} in a classical explanation, which somehow has not been successful. 
The literature on both experimental and theoretical fronts is so vast that we make reference to a recent review article \cite{Ba09}.
The failure of an explanation within  classical physics  has led to quantum mechanical topological explanation where the AB phase of the wave function of a charged particle depends on the
topology of the space it moves in, assuming that the presence of a solenoid of current makes the configuration space non-simply connected \cite{Ah83,Ah16,Ah17}. Such non-locality features of quantum mechanics may have deep philosophical implications \cite{16,He97}; classical explanations of EM momentum are not in vogue in the contemporary literature. However, as we shall demonstrate, without venturing into  topological questions, a description of this curious phenomenon through a viable explanation of the EM momentum is still possible in the way originally envisaged.
\section{Phase shift due to a solenoid of current between the two electron beams}
A long thin solenoid of current comprising circular loops of current, generates at a distance $R$ from the central axis of the solenoid, a magnetic field (in cgs units) given by \cite{PU85} 
 \begin{eqnarray}
\label{eq:84b.4b1a1}
\mathbf {B} &=&\frac{4\pi I M}{c} \hat z\,,\,\,\,\,\,\,\,\,\,\,\,\, R<r\\
\label{eq:84b.4b1a2}
\mathbf {B} &=&0 \,.\,\,\,\,\,\,\,\,\,\,\,\,\,\,\,\,\,\,\,\,\,\,\,\,\,\,\,\,\, R>r
\end{eqnarray}
where $I$ is the constant electric current flowing in the solenoid, with $M$ current loops per unit length along the $z$-axis, and $r$ is the radius of each circular loop. 
Thus there is no magnetic field outside the solenoid. 
%Also $\nabla \cdot {\mathbf A}=0$. 

Vector potential, defined at a field point by $\nabla \times \mathbf {A}=\mathbf {B}$, is easily determined for such a long solenoid of current, using Stokes' theorem, to get \cite{25}
\begin{eqnarray}
\label{eq:84a.4b2a}
\oint \mathbf {A} \cdot {\rm d}\mathbf {x}&=&\int( \nabla \times \mathbf {A}) \cdot {\rm d}{\mathbf a}=\int \mathbf {B} \cdot  {\rm d}{\mathbf a}\,,
\end{eqnarray}
which, from the cylindrical symmetry, possesses a finite component only along the azimuthal direction, with  ${\mathbf A}=A_\phi \hat\phi$, having a value at $R$ as
\begin{eqnarray}
\label{eq:84b.4b1a}
A_\phi &=&\frac {2\pi R I M}{c} \,\,\,\,\,\,\,\,\,\,\,\,\,\,\,\,\ R<r\\
\label{eq:84b.4b1b}
A_\phi &=&\frac {2\pi r^2 I M}{cR} \,\,\,\,\,\,\,\,\,\,\,\,\,\,\,\,\
 R>r
\end{eqnarray}

%In the AB double-slit experimental setup, a long but thin solenoid of electric current is introduced between the two slits (Fig.~1). , comprising circular loops of current,  the magnetic field (in cgs units) is \cite{PU85} 

%The magnetic field is determined from ${\mathbf A}=A_\phi \hat\phi$ as
%%Also $\nabla \cdot {\mathbf A}=0$. 
%\begin{eqnarray}
%\label{eq:84b.4b1}
%\mathbf {B} &=&\nabla \times \mathbf {A}=-\frac{\partial A_\phi}{\partial z} \hat R+\frac {1}{R}\frac {\partial (R A_\phi)}{\partial R} \hat z\,,
%\end{eqnarray}
% to get
%---------------------------------------------
\begin{figure}
\begin{center}
\includegraphics[width=8.9cm]{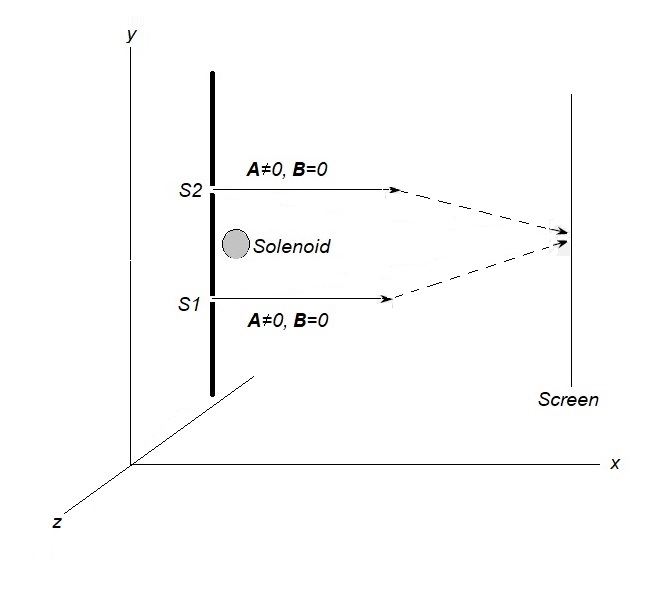}
\caption{A schematic of the AB experimental setup. Coherent beams of electrons, passing  through two slits, $S1$ and $S2$, separated along the y-axis, form an interference pattern on the screen. When a long but thin solenoid of electric current is introduced behind the two slits, midway between the two beams, an extra phase between the two electron beams appears, as shown by a shift in the interference pattern. In the region of electron beams, the magnetic field is nil ($\mathbf {B}=0$) though the vector potential is finite ($\mathbf {A}\ne 0$).}
\end{center}
\end{figure}

An electric charge $Q$, of mass $m$ and moving with a velocity ${\mathbf v}$, and accordingly having a mechanical momentum ${\mathbf p}_{\rm m}=m{\mathbf v}$ (we consider here only non-relativistic cases), when passing through a region with a finite vector potential $\mathbf A$, gets associated with it, in addition, an EM momentum   
\begin{eqnarray}
\label{eq:84a.3}
{{\mathbf p}_{e}}&=&\frac{Q{\mathbf A}}{c}\,.
\end{eqnarray}
This EM momentum is independent of either mass $m$ or velocity ${\mathbf v}$ of the charge $Q$, which may thus even be stationary. This form of momentum in classical  electromagnetism,  suggested by Maxwell \cite{Ma65}, has been discussed at some length in the literature \cite{Ko78,Se98,Gr12}. From Eq.~(\ref{eq:84a.3}) the momentum would be finite for a non-zero $\mathbf A$ at the location of $Q$ outside the solenoid of current, even if the magnetic field ${\displaystyle \mathbf {B} =\mathbf {0}}$ there. 

In quantum mechanics, the wave function associated with the electric charge $Q$, because of the additional EM momentum ${{\mathbf p}_{e}}$, develops an extra phase shift
\begin{eqnarray}
\label{eq:84a.4}
{\displaystyle \varphi=\frac{1}{\hbar}\int {\mathbf p}_{\rm e} \cdot {\rm d}{\mathbf x}=\frac{Q}{c\hbar}\int {\mathbf A} \cdot {\rm d}{\mathbf x}\,.}
\end{eqnarray}

In the AB experimental setup, a double-slit quantum mechanics experiment, a long but thin solenoid of current is introduced between the two coherent beams of electrons behind the slits (Fig.~1), 
%where every eigenfunction for $\mathbf A=0$ is multiplied by $e^{i\varphi }$.  
Now, two equal charges, $Q_1$  and $Q_2$, in the two coherent beams placed symmetrically on two opposite sides of the solenoid, and thus having equal and opposite vector potentials, ${\mathbf A}_1$, ${\mathbf A}_2$ with ${\mathbf A}_1=-{\mathbf A}_2$ at their locations, will accordingly, possess equal and opposite EM momentum vectors, $Q_1{\mathbf A}_1/c=-Q_2{\mathbf A}_2/c$.
%---------------------------------------------

Then the two charge beams moving along two different paths, but each having the same start and end points as the other, will acquire accordingly, a phase difference 
\begin{eqnarray}
\label{eq:84a.4a}
\Delta \varphi&=&\frac{Q}{c\hbar}\oint \mathbf {A} \cdot {\rm d}\mathbf {x}
%=\frac{Q}{c\hbar}\int( \nabla \times \mathbf {A}) \cdot {\rm d}{\mathbf a}\nonumber\\
=\frac{Q}{c\hbar}\int \mathbf {B} \cdot  {\rm d}{\mathbf a}
=\frac{Q\:\Phi}{c\hbar}\,,
\end{eqnarray}
which is thus determined by the magnetic flux $\Phi$ through the area enclosed between the paths, even though at the location of either beam the magnetic field is nil. The choice of gauge for ${\mathbf A}$ does not affect $\Delta \varphi$. This is because even though ${\mathbf A}$ may be arbitrary to the extent where gradient of some scalar function ($\nabla\psi$) may be added. However, while computing the extra phase difference (Eq.~(\ref{eq:84a.4a})) we integrate the tangential component of ${\mathbf A}$ around a closed path, but the integral of the tangential component of a gradient around a closed path is always zero \cite{29}. This can also be seen as the magnetic field $\bf B$ or the magnetic flux $\Phi$, and consequently $\Delta \varphi$ in Eq.~(\ref{eq:84a.4a}), do not depend upon the chosen gauge. Once Eq.~(\ref{eq:84a.3}), from classical  electromagnetism, is accepted to be providing a measure of EM momentum in the system, everything else seems to follow from quantum mechanics. 

Such a phase difference has actually been inferred in the AB experimental setup, from an observed shift in the interference pattern \cite{Ch60,To86,Os86}. 
From a physical perspective, however, it is not clear from where does such a mysterious EM momentum appear, whose presence has been verified experimentally, implying that there is some interaction between the solenoid and the charge beams even in the absence of any EM field at the locations of the beams. 
This interaction, ostensibly through the vector potential at the location of either beam, is responsible for EM momenta that get reflected in the observed quantum mechanical phase shift between two beams.
But to date it still remains a mystery how the solenoid influences the beams in the absence of  EM fields at their locations, or how do the beams interact with the solenoid.

\section{EM momentum of a charge in a vector potential from classical perspective}
%On the other hand, 
%after all 
Since Eq.~({\ref{eq:84a.3}), relating the momentum to the vector potential, has its genesis in classical electromagnetism, one should be able to resolve the issue of EM momentum, without invoking quantum effects, within the classical physics itself.
% which is our intention here. 
A convincing argument that $Q{\mathbf A}/c$ in some respects does represent momentum within the classical electromagnetism, comes from the following.

%Let us consider a case where electric current $I$ in the solenoid is slowly decreased at a constant rate. \cite{Jo94} From Eq.~({\ref{eq:84b.4b1a1}), $\mathbf {B}$ will be reducing too. Then Faraday's law of induction, $c\nabla \times \mathbf {E}=-\partial\mathbf {B}/\partial t$, and Stokes' theorem, give 
%\begin{eqnarray}
%\label{eq:84a.4b2c}
%\oint \mathbf {E} \cdot {\rm d}\mathbf {x}&=&\int( \nabla \times \mathbf {E}) \cdot {\rm d}{\mathbf a}=-\frac{1}{c}\int\frac{\partial \mathbf {B}}{\partial t}  \cdot  {\rm d}{\mathbf a}\,,
%\end{eqnarray}
%which, from cylindrical symmetry, yields ${\mathbf E}=E_\phi \hat\phi$, implying an azimuthal electric field outside the solenoid as
%\begin{eqnarray}
%\label{eq:84a.5}
%E_{\phi}&=&-\frac {\pi r^2 }{2\pi R}\frac{4\pi M}{c^2}\,\frac{{\rm d}{I}}{{\rm d}t}
%%=-\frac {2\pi r^2 M}{c^2R}\frac{{\rm d}{I}}{{\rm d}t}=-\frac{1}{c}\frac{{\rm d}{A_{\phi}}}{{\rm d}t}
%=-\frac{1}{c}\frac{\partial {A}_{\phi}}{\partial t}\,,
%\end{eqnarray}
%where we have used  Eq.~({\ref{eq:84b.4b1b}). We could have derived it directly from $c \mathbf {E}=-\partial\mathbf {A}/\partial t$, in the absence of a scalar potential. 
Suppose the electric current $I$ in the solenoid is slowly decreased at a constant rate, Then from Eq.~({\ref{eq:84b.4b1b}), the vector potential $\mathbf {A}$ at $R$, location of $Q$, would decrease too,  
giving rise, in the absence of a scalar potential, to an electric field $c \mathbf {E}=-\partial\mathbf {A}/\partial t$.
This electric field would exert a force on the stationary $Q$, changing its mechanical momentum
\begin{eqnarray}
\label{eq:84a.5a}
\frac{{\rm d}{\mathbf p}_{\rm m}}{{\rm d}t}&=&m\frac{{\rm d}{\mathbf v}}{{\rm d}t}=Q\mathbf {E}\nonumber\\
&=&-\frac{Q}{c}\frac{{\rm d}{\mathbf A}}{{\rm d}t}
=-\frac {2\pi Q r^2 M}{c^2R}\frac{{\rm d}{I}}{{\rm d}t} \hat\phi
\,,
\end{eqnarray}
implying thereby that the mechanical momentum $m{\mathbf v}$ of $Q$ increases at the cost of $Q{\mathbf A}/{c}$, with the latter  decreasing for ${\rm d}{I}/{\rm d}t <0$.  
This suggests
%highly suggestive for 
$Q{\mathbf A}/c$ to be a form of momentum, called EM momentum ${\mathbf p}_{\rm e}$ of the charge $Q$, along with conservation of ${\mathbf p}_{\rm m}+{\mathbf p}_{\rm e}=m{\mathbf v}+{Q{\mathbf A}}/{c}$, called generalized momentum \cite{Go50}, as 
%${\mathbf p}_{\rm m}+{\mathbf p}_{\rm e}$
\begin{eqnarray}
\label{eq:84a.6}
\frac{\rm d}{{\rm d}t}\left[{\mathbf p}_{\rm m}+{\mathbf p}_{\rm e}\right]&=& 0\,,
\end{eqnarray}
%along with the conservation of generalized momentum, ${\mathbf p}_{\rm m}+{\mathbf p}_{\rm e}$, 
all within classical electromagnetism. That the conservation of EM plus kinetic momentum is of interest in AB effect has recently been emphasized \cite{Es23}.
%
%
%While a time derivative of $Q{\mathbf A}/c$ does seem to represent a temporal rate of change of momentum of the system in classical electromagnetism (even when the magnetic field might be still zero at the location of $Q$), however, it does not explain what gives rise to the EM momentum in a steady state of the system. Since the charge $Q$ lying outside the solenoid experiences no EM fields in the absence of any temporal variations of $I$, ${\mathbf A}$ and ${\mathbf B}$, a steady EM momentum $Q{\mathbf A}/c$ in the system, still remains to date an unresolved enigma. 

However, to date, it still remains an unresolved enigma where after all could such a `mysteriously hidden' momentum be residing since there is no obvious net linear motion in the system, especially when the charge is considered to be stationary. 
As the term ``momentum'' conjures up a vision of some kind of linear motion a question arises where, after all such motion, if any, is lying in the system? The only non-random motion ostensibly present in the system is in the drift velocities of the current carrying charges in the steady current loop. However, a linear momentum cannot be solely due to the drift velocities, as any such momentum vector integrated over a closed circuit would be zero. Moreover, from Eq.~(\ref{eq:84a.3}), the momentum in question involves, not just the electric current that gives rise to ${\mathbf A}$, but also the specification of charge $Q$ along with {\em its location} (where ${\mathbf A}$ is to be evaluated), although any movement of $Q$ does not enter into picture.

We explore here accordingly, from a classical physics perspective, the mysterious momentum in the AB setup, endeavouring to possibly unravel wherein the momentum lies in the system. We shall demonstrate here explicitly how an electric charge stationary at a location, where there may be a finite vector potential ${\mathbf A}$, though no magnetic field, does give rise to EM momentum, mysteriously latent in the system and which is reflected in the AB experiments.  
Moreover, as will be seen, the momentum in question is not confined to and localized at some specific location, like that of charge $Q$; the non-local characteristic of momentum is evident  in such a case even within the classical physics picture itself.
%%%%%%%%%%%%%%%%%%%%%%%%%%%%%%%%%%%%%%%%%%

In the case of $N$ discrete charges $q_{\rm j}$, with velocity vectors ${\mathbf v}_{\rm j}$, the vector potential ${\mathbf A}$ at the  location of the charge $Q$ is computed from the summation
\begin{eqnarray}
\label{eq:84a.4e1}
{\mathbf A}({\mathbf x}_{0})&=&\frac {1}{c}\sum_{j=1}^{N} \frac {q_{\rm j}\,{\mathbf v}_{\rm j}}{|{\mathbf x}_{\rm j}-{\mathbf x}_{0}|}\,,
\end{eqnarray}
where $|{\mathbf x}_{\rm j}-{\mathbf x}_{0}|$ is the distance of charge $q_{\rm j}$, moving with velocity ${\mathbf v}_{\rm j}$, from the location ${\mathbf x}_{0}$ of the charge $Q$. 

Then we have  
\begin{eqnarray}
\label{eq:84a.4e2}
\frac {Q {\mathbf A}}{c}&=&\frac {1}{c^2}\sum_{j=1}^{N} \frac {Q\, q_{\rm j}}{|{\mathbf x}_{\rm j}-{\mathbf x}_{0}|}\,{\mathbf v}_{\rm j}\,.
\end{eqnarray}

In order to comprehend the EM momentum in the system from a physical perspective, instead of the usual way of looking at it in terms of the vector potential ${\mathbf A}$ at the location ${\mathbf x}_0$ of the charge $Q$, we can interpret Eq.~(\ref{eq:84a.4e2}) in terms of the scalar potential $\phi_{\rm j}=Q/|{\mathbf x}_{\rm j}-{\mathbf x}_{0}|$, due to $Q$ at the location ${\mathbf x}_{\rm j}$ of charge $q_{\rm j}$. 
For that we rewrite Eq.(\ref{eq:84a.4e2}) as 
\begin{eqnarray}
\label{eq:84a.4e3}
\frac {Q {\mathbf A}}{c}&=&\sum_{j=1}^{N} \frac {\phi_{\rm j}\:q_{\rm j}}{c^2}\,{\mathbf v}_{\rm j}%=\sum_{j=1}^{N} {m_{\rm j}\,{\mathbf v}_{\rm j}}
\,.
\end{eqnarray}
%%---------------------------------------------

We can use the energy-mass relation, to express the potential energy $q_{\rm j}\,\phi_{\rm j}$ of a charge $q_{\rm j}$, owing to the presence of charge $Q$ at ${\mathbf x}_0$, in terms of its mass equivalent
\begin{eqnarray}
\label{eq:84a.4d4}
\Delta m_{\rm j}&=&\frac {q_{\rm j}\,\phi_{\rm j}}{c^2} =\frac {1}{c^2} \frac {Q\, q_{\rm j}}{|{\mathbf x}_{\rm j}-{\mathbf x}_0|}\,.
\end{eqnarray}
Then we can write
\begin{eqnarray}
\label{eq:84a.4e4}
\frac {Q {\mathbf A}}{c}&=&\sum_{j=1}^{N} {\Delta m_{\rm j}\,{\mathbf v}_{\rm j}}
%={{\mathbf p}_{e}}
\,,
\end{eqnarray}
which can now be readily recognized as an EM momentum, ${\mathbf p}_{e}$, in the system. 
It should be noted that $\Delta m_{\rm j}$ here has nothing to do with mass $m_{\rm j}$ of the $j$th charged particle and that the EM momentum in Eq.~(\ref{eq:84a.4e4}) is not sum of the kinetic momentum, $\Sigma m_{\rm j}\,{\mathbf v}_{\rm j}$, of moving charged particles. Also it is not possible to localize the potential energy $q_{\rm j}\,\phi_{\rm j}$ or the equivalent mass $\Delta m_{\rm j}$ at either of the charge locations, ${\mathbf x}_{0}$ or ${\mathbf x}_{\rm j}$. Nor could one pinpoint the EM momentum ${\mathbf p}_{e}$ at the location ${\mathbf x}_{0}$ of charge $Q$, one has to instead take a holistic view that the system comprises an EM momentum ${\mathbf p}_{e}$, without localizing it, even from a classical physics perspective.

%%%%%%%%%%%%%%%%%%%%%%%%%%%%%%%%%%%%%%%%%%%%%%%%%%%%%%%%%%%%%%%%%%%
For a continuous distribution of moving charges or a current density ${\mathbf j}({\mathbf x})=\rho({\mathbf x})\,{\mathbf v}({\mathbf x})$, the vector potential $\mathbf A$ at a field point ${\mathbf x}_0$ is determined from the volume integral  \cite{PU85,25,1,2}
\begin{eqnarray}
\label{eq:84a.4b}
{\mathbf A}({\mathbf x}_0)&=&\int\frac {{\mathbf j}({\mathbf x})}{c|{\mathbf x}-{\mathbf x}_0|}\,{\rm d}\tau=\frac {1}{c}\int\frac {\rho({\mathbf x})\,{\mathbf v}({\mathbf x})}{|{\mathbf x}-{\mathbf x}_0|}\,{\rm d}\tau\,,
\end{eqnarray}
where $|{\mathbf x}-{\mathbf x}_0|$ is the distance from the point ${\mathbf x}_0$ of the charge element $\rho({\mathbf x})\,{\rm d}\tau$, moving with a velocity $\mathbf v({\mathbf x})$, here ${\rm d}\tau$ denotes an element of volume. 
%We assume all motions to be non-relativistic.

Then we can write
\begin{eqnarray}
\label{eq:84a.4d1}
\frac{Q{\mathbf A}}{c}=\frac {1}{c^2}\int\frac {Q\,\rho({\mathbf x})\,{\mathbf v}({\mathbf x})}{|{\mathbf x}-{\mathbf x}_0|}\,{\rm d}\tau\,,
\end{eqnarray}
%In order to comprehend from a physical perspective the presence of EM momentum  in Eq.~(\ref{eq:84a.4d1}), we can look at it in an alternate way. Due to scalar potential $\rho\,{\rm d}\tau/|{\mathbf x}-{\mathbf x}_{0}|$ of the charge element  $\rho\,{\rm d}\tau$, the charge $Q$ at the location  ${\mathbf x}_{0}$ has a potential energy  $Q\rho\,{\rm d}\tau/|{\mathbf x}-{\mathbf x}_{0}|$. 
Because of the scalar potential $\phi({\mathbf x})=Q/|{\mathbf x}-{\mathbf x}_{0}|$ at ${\mathbf x}$ due to $Q$, the system comprising a charge density $\rho({\mathbf x})$, possesses $\rho({\mathbf x})\,\phi({\mathbf x})$ as potential energy per unit volume. Then in the expression  
%Eq.~(\ref{eq:84a.4d1}) as
\begin{eqnarray}
\label{eq:84a.4c1}
\frac{Q{\mathbf A}}{c}&=&\frac {1}{c^2}\int {\phi({\mathbf x})\,\rho({\mathbf x})\,{\mathbf v}({\mathbf x})}\,{\rm d}\tau\,,
\end{eqnarray}
%Thus the system comprising a charge density $\rho({\mathbf x})$, possesses $\rho({\mathbf x})\,\phi({\mathbf x})$ as potential energy per unit volume, owing to the presence of charge $Q$ at ${\mathbf x}_0$. 
we could use the energy-mass relation, to express the potential energy density in terms of its equivalent mass density, $\mu({\mathbf x})={\rho({\mathbf x})\phi({\mathbf x})}/{c^2}$, to write     
%\begin{eqnarray}
%\label{eq:84a.4d5}
%\mu({\mathbf x})&=&\frac{\rho({\mathbf x})\phi({\mathbf x})}{c^2} \,,
%\end{eqnarray}
%to get
\begin{eqnarray}
\label{eq:84a.4d2}
\frac{Q{\mathbf A}}{c}&=&\int{\mu({\mathbf x})\,{\mathbf v}({\mathbf x})}\,{\rm d}\tau={{\mathbf p}_{e}}\,.
\end{eqnarray}
Here $\mu({\mathbf x}){\mathbf v}({\mathbf x})$ is the momentum density in the system,
whose volume integral yields the EM momentum ${{\mathbf p}_{\rm e}}$ in the system, owing to the influence of charge $Q$, lying at ${\mathbf x}_0$, on the charges moving in the volume.
%\begin{eqnarray}
%\label{eq:84a.4d3}
%\frac{Q{\mathbf A}}{c}&=&{{\mathbf p}_{e}}\,,
%\end{eqnarray}
\section{EM momentum of a charge outside a solenoid of current}
We want now to examine the EM momentum of a charge outside a solenoid and we shall show here that Eqs.~(\ref{eq:84a.4d1}), (\ref{eq:84a.4c1}), or equivalently (\ref{eq:84a.4d2}), lead to an EM field momentum, latent in the system. We should clarify that this EM momentum is different from what sometimes is referred to as `hidden momentum' \cite{Sh67}
%%%%%%%%%%%%%%%%%%%%%%%%%%%%%%%%%%%%%%%%%%%%%%%%%%%%%%%%%%%%%%%%%%%%%%%%%  suppressed temporarily
and which actually is a mechanical momentum \cite{31}, present in the system.
%%%%%%%%%%%%%%%%%%%%%%%%%%%%%%%%%%%%%%%%%%%%%%%%%%%%%%%%%%%%%%%%%%%%%%%%% end of temporarily  suppressed 

A long, thin solenoid, carrying a steady electric current $I$, can be considered as a superposition of large number of small planar current loops, each carrying current $I$, with, say, $M$ loops stacked per unit length of the solenoid, plus a current $I$ along the $z$-axis on the solenoid surface. For the external field, the surface current along the $z$-axis is equivalent to a long straight wire carrying current $I$ along the axis of the solenoid. Inside the solenoid, the magnetic field would still be $B_{\rm z}=4 \pi I M/c$ (Eq.~({\ref{eq:84b.4b1a1})). 
On the outside, however, there will be an azimuthal field $B_{\phi}= {2I}/{cR}$ due to current $I$ along the axis of the solenoid \cite{PU85}, which nonetheless, would not affect $\Delta \varphi$ (Eq.~({\ref{eq:84a.4a})), since flux $\Phi$ enclosed between the two beams will not change. Therefore we shall henceforth ignore the axial current in the solenoid.
%In case, one brings down the current along the axis of the solenoid, this extra $A_{\rm z}$ and $B_{\phi}$ will disappear.is equivalent to a superposition of large number, each with an electric current $I$, stacked one above the other, and a linear current, $I$, flowing.  

A small planar loop carrying a current $I$ around area $\mathbf a$ of the loop constitutes, irrespective of its shape, a magnetic dipole ${\mathbf m}=I{\mathbf a}/c$ \cite{PU85}. The vector potential, ${\bf A}$  from a magnetic dipole $\bf m$ at a radius vector ${\mathbf R}$ from the dipole is given as \cite{PU85}
\begin{eqnarray}
\label{eq:84b.4}
{\mathbf A}=\frac{{\mathbf m}\times\hat{\mathbf R}}{R^2}\,,
\end{eqnarray}
This implies for the charge $Q$ at ${\mathbf R}$, from Eq.~({\ref{eq:84a.3}), an EM momentum 
\begin{eqnarray}
\label{eq:100.4}
{\mathbf p}_{\rm e} &=&\frac{Q\,{\mathbf A}}{c}=\frac{Q\,{\mathbf m}\times\hat{\mathbf R}}{cR^2}=\frac {({\mathbf E} \times {\mathbf m})}{c}\nonumber\\
&=&\frac {({\mathbf E} \times {\mathbf a})\,I}{c^2},
\end{eqnarray}
where ${\mathbf E}=-Q\hat{\mathbf R}/{R^2}$ is the electric field of the charge $Q$ at the location of the small current loop. Thus Eq.~({\ref{eq:84a.3}), expressing EM momentum due to the vector potential of the current loop at the location of $Q$, represents implicitly a mutual interaction of the charge $Q$ and the current loop since from Eq.~({\ref{eq:100.4}), the EM momentum vector could as well be considered due to the cross product of electric field ${\mathbf E}$ of $Q$ and magnetic moment ${\mathbf m}$ of the current loop. Of course this does not in any way further clarify the question of momentum in this apparently static system.

Since the current loop may consist of a conducting wire, the electric field of the charge $Q$ will not extend inside the loop wire, as the induced surface charge density there would tend to cancel any external static electric field inside the wire, leaving only the perpendicular components at the surface, to make the loop equipotential. In that case the system, in the absence of mutual interaction of charge $Q$ and the current loop, may not possess an EM momentum \cite{Jo94}, contrary to what would have been otherwise expected from Eq.~({\ref{eq:84a.3}). However, the EM momentum inferred for the system from the experimentally  observed shift in the fringe patterns \cite{Ch60,To86,Os86} indicates that there may be  something amiss in the above arguments.

Actually, there is a rather subtle issue involved here as in these experiments there are two coherent charged beams, emerging simultaneously from slits $S1$ and $S2$, assumed to be symmetrically placed on either side of the thin solenoid or its small current loops (Fig.~1). Thus one has to consider the EM momentum simultaneously for a pair of equal charges, placed symmetrically, on two opposite sides of the current loop. To be specific, we designate the charges as  $Q_1$ from slit $S1$ and $Q_2$ from $S2$, lying respectively at distances ${\mathbf R}_1$ and ${\mathbf R}_2$, with ${\mathbf R}_1=-{\mathbf R}_2$, measured from the loop position.
Thus, at least to a first order, the electric fields, say ${\mathbf E}_1$ and  ${\mathbf E}_2$ at the loop location and the corresponding scalar potentials across the small loop will be equal and opposite for the two charges, making the loop effectively equipotential, even without the aid of induced surface charges that would have otherwise got formed there. 

One might object that the assumed symmetric situation, where interfering electrons emerge from the two slits simultaneously may not always hold good as the interference patterns on the screen arise even if there is only one electron in the system at a time, and that assumption of electrons emerging from both slits 'simultaneously' may not be always justified. Here it may be pointed out that an electron does not go through either slit 1 or slit 2, and that the only way to generate an interference pattern is if the  electron simultaneously goes through both slits \cite{Fe65,Wi71}.

Since the induced surface charge on the conducting coil due to either charge may be nil, the mutual interaction between the current loop and each charge would still be present, with the resulting momentum associated with either charge being equal and opposite to that associated with the other charge, 
This can be seen from the conservation of generalized momentum. Suppose the electric current $I$ in the solenoid is slowly reduced at some constant rate. As we discussed earlier, the induced electric field vector (Eq.~({\ref{eq:84a.5a}))
at ${\mathbf R}_1$ would then exert force on $Q_1$, changing its mechanical momentum. Now  any such change in the mechanical momentum of a charge is possible, from the conservation of generalized momentum (Eq.~({\ref{eq:84a.6})), only at the cost of its EM momentum, $\Delta {\mathbf p}_{\rm m1}=-\Delta {\mathbf p}_{\rm e1}$, implying the presence of ${{\mathbf p}_{e}}_1$ in the system. Similar is the argument for ${{\mathbf p}_{e}}_2$ in the case of $Q_2$. 

Thus one could proceed with the investigation of EM momentum associated with either charge, without considering any induced surface charges in the current loop that otherwise might have formed to cancel the electric fields, thus leaving the mutual interaction between corresponding charge and the current loop unperturbed.
%---------------------------------------------
\begin{figure*}
\begin{center}
\includegraphics[width=12cm]{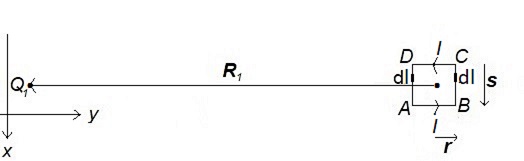}
\caption{A charge $Q_1$ lies a distance $R_1$ from a small current loop $ABCD$, carrying a uniform current $I$. Two equal current elements, ${\rm d}{l}$,
separated by distance $2r$, on two opposite sides of the current loop, are shown.}
\end{center}
\end{figure*}
%%%%%%%%%%%%%%%%%%%%%%%%%%%%%%%%%%%%%%%%%%%%%%%%%%%%%%%%%%%%%%%%%%%%%%%%
%%%%%%%%%%%%%%%%%%%%%%%%%%%%%%%%%%%%%%%%%%%%%%%%%%%%%%%%%%%%%%%%%%%%%%%%
\subsection{EM momentum of a charge outside a small current loop}
We apply our above results to show, from a physical perspective, that in the presence of an external charge, a current loop does give rise to an EM momentum. 
For this, we consider a small rectangular loop, carrying a steady current, $I=j \sigma=n e  v_{\rm d} \sigma $, where $n$ is the number density of charges in the circuit, 
$e$ is the electric charge of conducting charged particles, $v_{\rm d}$ is their drift velocity and $\sigma$ is the cross-section of the current-carrying wire. Then from Eq.~(\ref{eq:84a.4e1}) or (\ref{eq:84a.4b}), the vector potential $\mathbf A$ at the location ${\mathbf x}_0$ of $Q$ is determined from the integral over the circuit length
\begin{eqnarray}
\label{eq:84a.4b2b}
{\mathbf A}({\mathbf x})&=&\frac {1}{c}\oint\frac {n e \sigma}{|{\mathbf x}-{\mathbf x}_0|}{\mathbf v}_{\rm d}\,{\rm d}{l}\,,
\end{eqnarray}
where ${\rm d}{l}$ is an element of  the circuit, with ${\mathbf v}_{\rm d}$ as the drift velocity along  the direction of the current at the location ${\mathbf x}$ of the circuit.  
Then the EM momentum of the system, from Eq.~(\ref{eq:84a.4d1}), can be written as 
\begin{eqnarray}
\label{eq:84a.4c}
{{\mathbf p}_{e}}&=&\oint\frac{{Q}\,{n e \sigma}{\mathbf v}_{\rm d}}
{ {c^2}|{\mathbf x}-{\mathbf x}_0|}{\rm d}{l}=\oint\frac{{\mathbf v}_{\rm d}}{c^2}{\rm d}\,{\cal E}=\oint{\mathbf v}_{\rm d}\,{\rm d}{m}.
\end{eqnarray}
Here ${\rm d}{\cal E}={Q}\,{n e \sigma}{\rm d}{l}/|{\mathbf x}-{\mathbf x}_0|$ is the electric potential energy of current carrying charges in the volume element $\sigma{\rm d}{l}$ in the presence of charge $Q$ at ${\mathbf x}_0$, while  ${\rm d} m={\rm d}{\cal E}/(c^2)$ is the equivalent electric mass element. The momentum in Eq.~(\ref{eq:84a.4c}) is not the same as the kinetic momentum, $\oint m n \sigma {\mathbf v}_{\rm d}\,{\rm d}{l}$, of the electric current carriers, each supposedly of individual mass $m$. Such a kinetic momentum of the electric current carriers for a steady current, in any case, yields a nil value when summed over the whole current loop ($ m n \sigma \oint{\mathbf v}_{\rm d}\,{\rm d}{l}=0$), however, $\oint {\mathbf v}_{\rm d}\,{\rm d}{m}$ over the loop does result in a finite value, as will be shown below.

%---------------------------------------------
\subsection{EM momentum of charges placed symmetrically on opposite sides a small current loop}
We have still to show that the pair of equal charges, $Q_1$ and $Q_2$, placed symmetrically, on two opposite sides of the current carrying solenoid will give rise to equal and opposite EM momentum in the system, something not so readily apparent from Eqs.~(\ref{eq:84a.4e4}), (\ref{eq:84a.4d2}) or (\ref{eq:84a.4c}). 

Let us consider charge $Q_1$ at a distance $R_1$ along the $-y$ direction, from a small rectangular current loop (Fig.~2). We can calculate the EM momentum of the system by considering pairs of current elements, each of infinitesimal length ${\rm d}{l}$, placed symmetrically on two opposite sides of the loop. 
The drift velocity in the arm $DA$ of the current loop is along $\hat{\mathbf x}$ direction and the current element is nearer to the charge $Q_1$, thus making a higher contribution to the EM momentum which is along $\hat{\mathbf x}$ direction, while the drift velocity in the arm $BC$ is along $-\hat{\mathbf x}$ direction and the current element being farther from $Q_1$, makes a contribution which is relatively lower and is along $-\hat{\mathbf x}$ direction. The EM momentum of the whole  system, for a small current loop, with $r\ll R_1$ (Fig.~2), is 
\begin{eqnarray}
\label{eq:100.1}
{\mathbf p}_{\rm e1} &=& \frac{{Q_1}\,{n e \sigma}\,s\,{v}_{\rm d}}{{c^2}(R_1-r)} \hat{\mathbf x}-\frac{{Q_1}\,{n e \sigma}\,s\,{v}_{\rm d}}{{c^2}(R_1+r)} \hat{\mathbf x}
%=\frac{{Q_1}\,I\,s \,r}{2{c^2}R_1^2} \hat{\mathbf x}
\,.
\end{eqnarray}
The first term on the right hand side is the contribution of the current loop from the arm DA  while the second term is from the arm BC, each arm of length $s$. 
The contributions from the arms $AB$ and $CD$ to the EM momentum, being equal and opposite, cancel. 

To a first order (for $r/ R_1\ll 1$), we get
\begin{eqnarray}
\label{eq:100.2}
{\mathbf p}_{\rm e1} &=& \frac{{2Q_1}\,{n e \sigma}{v}_{\rm d}\,sr}{{c^2}R_1^2} \hat{\mathbf x}%=\frac{2{Q_1}\,I\,s \,r}{{c^2}R_1^2} \hat{\mathbf x}
=\frac {({\mathbf E}_1 \times {\mathbf a})\,I}{c^2}=\frac {({\mathbf E}_1 \times {\mathbf m})}{c}\nonumber\\
&=&\frac{Q_1\,{\mathbf m}\times\hat{\mathbf R}_1}{cR_1^2}=\frac{Q_1\,{\mathbf A}_1}{c}\,,
\end{eqnarray}
where ${\mathbf E}_1=Q_1{\hat{\mathbf y}}/{R_1^2}$, is the electric field of the charge $Q_1$ at the location of the small current loop, ${\mathbf a}=2\:{\mathbf s}\times {\mathbf r}=2s\,r\, \hat{\mathbf z}$ is the area vector of the current loop and ${\mathbf A}_1$ is the vector potential at the location of $Q_1$.
The EM momentum in the system, being directly proportional to the drift velocity (${p}_{\rm e}\propto {v}_{\rm d}$) of current carriers, is thus zero to start with for ${v}_{\rm d}=0$ (when $I=0$) and increases as ${v}_{\rm d}$ increases with $I$.
It is, however, interesting that a finite linear momentum exists in the system owing  to the presence of charge $Q_1$ outside the current loop, even when $Q_1$ as well as the current loop are both stationary, and momentum vector, if any, due to current carrying charges adds to zero over the closed loop ($ m n \sigma \oint{\mathbf v}_{\rm d}{\rm d}{l}=0$). 

As for the charge $Q_2$, lying on opposite side of the current loop at distance $R_2$ along the $y$ direction, the contribution of EM momentum from the arm BC of the current loop will be higher than that of the arm DA, as a result the net EM momentum will be along $-x$ direction. It will be so even though the potential energy and the mass equivalent for  $Q_2$ is similar as for  $Q_1$, the opposite directions of drift velocities in arms  DA and BC will make ${\mathbf p}_{\rm e2}=-{\mathbf p}_{\rm e1}$.

The electric current in the loop could be due to electrons instead of positive charges, essentially making no difference to any of our arguments. Also, at the locations of charge $Q_1$ or $Q_2$, the electric field as well as the scalar potential of the lattice of positive ions, is equal and opposite to that of negative current carrying electrons for an over all charge neutral current loop, however, that does not cancel the EM momenta ${{\mathbf p}_{e1}}$ and ${{\mathbf p}_{e2}}$, arising from the drift velocities of current carrier electrons as the positive ions, fixed in the lattice, do not move with any drift velocity. Net electric potential energy of current carrier electrons as well as of the positive ions together in the system is zero due to either $Q_1$ or $Q_2$, nonetheless a net EM momentum does exist in the system due to $Q_1$ and $Q_2$. 
 
We could replace the rectangular shape of the loop with a polygon comprising a larger number of sides or in the limit even with a circular loop without changing the final result. In fact, from a side by side superposition of a sufficiently large number of small rectangular loops, any such shape of the current loop could be realized. Further, the electric field $\mathbf E$ now need not necessarily be only in the plane of the current loop. Equation~(\ref{eq:100.2}), therefore, is a fairly general result as long as the current loop is small enough for $\mathbf E$ to be considered uniform over its extent. 
\subsection{An alternate computation of EM momentum for current loop in the electric field of a charge}
The presence of EM momentum (Eq.~(\ref{eq:100.2}) in the system, can be worked out in an alternate way, without starting from the potential energy of the current-carrying charges in the current loop, due to the charge $Q$ (Eq.~(\ref{eq:84a.4d1})). We could instead employ the force due to electric field of $Q$ on the current-carrying charges in the current loop. In fact, along with the computation of the EM momentum, one could even derive Eq.~(\ref{eq:84a.3}) as well, that way. 
\begin{figure}
\begin{center}
\includegraphics[width=8cm]{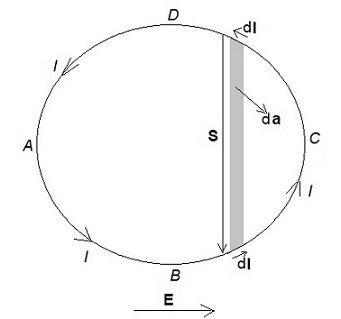}
\caption{A current loop $ABCD$ is carrying current $I$. There is an electric field ${\mathbf E}$, uniform over the current loop. The electric field ${\mathbf E}$ does positive  work, $I {\mathbf E} \cdot {\rm d}{\mathbf l}$ on a current element ${\rm d}{\mathbf l}$ in the arc $ABC$ of the current loop and an equal and opposite work on a similar current element ${\rm d}{\mathbf l}$ in the arc $CDA$, the two current elements separated by a distance $S$, on two opposite sides of the current loop. This implies transfer of an infinitesimal EM energy, between two current elements across the shaded area ${\rm d}{\mathbf a}$ of the circuit, which represents an element of EM momentum in the system.}
\end{center}
\end{figure}

We consider a small circular loop, $ABCD$, carrying a constant current $I=j \sigma$, with $\mathbf {j}$ as the current density and $\sigma$ the cross-section of the current-carrying wire. There is an electric field $\mathbf E$, uniform over the dimensions of the current loop (Fig.~3). 
The electric field does work on the current density ${\mathbf j}$ at a rate ${\mathbf j}\cdot {\mathbf E}$ per unit volume \cite{25,1,2}. 
Accordingly, on a current element d$\mathbf l$ of the loop, the electric field does work
at a rate ${\mathbf j}\cdot {\mathbf E}\,\sigma\, {\rm d}{l}= I\, {\rm d}{\mathbf l \cdot E}$.  
In section $ABC$ of the loop,  the electric field
$\mathbf E$ is doing positive work while in section $CDA$, the work done is negative. 

Thus, power is fed through the electric field into the system through the side $ABC$ while a similar amount of power is being drained off from the side $CDA$.  Effectively, a continuous transfer of electric energy per unit time is taking place from arm $CDA$ (where there is a loss of energy per unit time) to $ABC$ (where there is a gain in energy at the same temporal rate)  across the current loop due to the presence of the electrical field (Fig.~3). Even Though in this process there is no net
change in the total energy content of the system, there is nevertheless an  energy flux across the loop from side $CDA$ to $ABC$ and consequently, a linear momentum associated with this energy flux.  

The EM momentum can be calculated easily if we consider a pair of current
elements placed symmetrically on two opposite sides of the loop (Fig.~3). The power being fed into the circuit element ${\rm d}{\mathbf l}$ in the side $ABC$ 
through the electric field is $I {\mathbf E} \cdot {\rm d}{\mathbf l}$, while a similar amount of power is being drained off from a similar circuit element ${\rm d}{\mathbf l}$ in the side $CDA$. Effectively, an energy 
$I {\mathbf E} \cdot {\rm d}{\mathbf l}$ is being transported per unit time along $\textbf{s}$, the vector joining the  current element in side $CDA$ to that in $ABC$, implying a momentum
\begin{equation}
\label{eq:p31.1a}
{\rm d}{\mathbf  P}=I({\mathbf E} \cdot {\rm d}{\mathbf l}) \frac {\mathbf s}{c^2}= {\mathbf E} \times {\rm d}{\mathbf a}\frac {I}{c^2}, 
\end{equation}
where d$\mathbf a={\mathbf s}\times {\rm d}{\mathbf l}$ is the element of area vector contained between these two current elements across the loop. 
An integration over the entire loop, using Eq.~(\ref{eq:84b.4}), gives the total momentum 
\begin{equation}
\label{eq:p31.1b}
{\mathbf  P}_{\rm e}=\frac {({\mathbf E} \times {\mathbf a})\,I}{c^2}= \frac{\mathbf E \times \mathbf m}{c}=\frac{Q{\mathbf m}\times\hat{\mathbf R}}{R^2c}=\frac{Q {\mathbf A}}{c}\,, 
\end{equation}
%where ${\mathbf m=a}\,I/c$ is the magnetic dipole moment of the current loop. 
Thus in this way we get not only the EM momentum, which is the same as in Eq.~(\ref{eq:100.2}), we also derive Eq.~(\ref{eq:84a.3}).

%%%%%%%%%%%%%%%%%%%%%%%%%%%%%%%%%%%%%%%%%%%%%%%%%%%%%%%%%%%%%%%%%%%%%%%%%  suppressed temporarily
There are other, similar examples in physics where momentum is present in the system due to equal and opposite work being done on spatially separated parts of the system, implying a flux of energy between these spatially separated parts, implying momentum in the system. 
For instance, a similar continuous transport of EM energy per unit time across an electric circuit between its opposite arms, but with no change in the net energy of the system, has been shown elsewhere \cite{31} to explain the presence of linear EM momentum in a stationary system comprising a pair of crossed electric and magnetic dipoles, where nothing obviously is moving or no temporal changes are occurring in the system.
A perfect fluid under pressure, having a bulk motion even with non-relativistic velocities has finite momentum proportional to pressure that does work on two opposite ends of a fluid element giving rise to momentum in the system \cite{84}. There is also an opposite example where a charged parallel plate capacitor, moving parallel to the plate separation, has finite EM energy, but in spite of its motion, has zero EM momentum in the system \cite{15,84a}. 
A moving spherical charge distribution, representing a classical electron model, where an equal and opposite work done by the opposite forces of the leading and trailing hemispheres, gives rise to an energy flux and thereby an EM momentum  in the system that explains the more than a century-old famous factor of $4/3$ in the EM momentum of such a system \cite{15,84a}.
%%%%%%%%%%%%%%%%%%%%%%%%%%%%%%%%%%%%%%%%%%%%%%%%%%%%%%%%%%%%%%%%%%%%%%%%% end of temporarily  

One would normally expect the power difference between arms $ABC$ and $CDA$ to be compensated by the agency tending to maintain a uniform and steady electric current in the loop, with an equal amount of mechanical energy transfer rate from arm $ABC$ to $CDA$ in the circuit that itself might  entail a mechanical momentum  
%%%%%%%%%%%%%%%%%%%%%%%%%%%%%%%%%%%%%%%%%%%%%%%%%%%%%%%%%%%%%%%%%%%%%%%%%  suppressed temporarily
\cite{31}, 
%%%%%%%%%%%%%%%%%%%%%%%%%%%%%%%%%%%%%%%%%%%%%%%%%%%%%%%%%%%%%%%%%%%%%%%%% end of temporarily  
However, in the present case, with equal charges, $Q_1$ and $Q_2$, on opposite sides of the current loop, a constant current in both arms will be maintained because of their equal and opposite electric fields, ${\mathbf E}_1$ and ${\mathbf E}_2$. Moreover, the energy flux from arm $CDA$ to $ABC$ because of $Q_1$ will be compensated by an equal energy flux from arm $ABC$ to $CDA$ due to $Q_2$. However, as was discussed earlier, associated with individual charges there would still be present EM momenta, ${{\mathbf p}_{e}}_1$, ${{\mathbf p}_{e}}_2$, source of the phase difference, $\Delta \varphi$, experimentally observed between the two charge beams.

Now, we can compute from Eq.~(\ref{eq:100.2}) the total EM linear momentum associated with the charge $Q_1$ and the solenoid of current by summing over $M l$ current loops of the solenoid, and using ${B}_{\rm z}= {4\pi I M}/{c}$ inside the solenoid (Eq.~(\ref{eq:84b.4b1a1})), to get 
\begin{eqnarray}
\label{eq:84b.6}
{\mathbf p}_{\rm e1}=\sum_{1}^{Ml} \frac {({\mathbf E}_1 \times {\mathbf a})\,I}{c^2}=\int\frac {({\mathbf E}_1 \times {\mathbf B})}{4\pi c} {\rm d}\tau\,,
\end{eqnarray}
which is the volume integral of the EM momentum density, $({\mathbf E} \times {\mathbf B})/4\pi c$ over the solenoid. 
For the charge $Q_2$, with ${\mathbf E}_2=-{\mathbf E}_1$, Eq.~(\ref{eq:84b.6}) implies ${\mathbf p}_{\rm e2}=-{\mathbf p}_{\rm e1}$.

The seat of the field momentum (Eq.~(\ref{eq:84b.6})) might appear to be within the solenoid, however, one has to take the holistic view that the EM momentum actually lies in the {\em composite system} of the charge plus solenoid, as it has been emphasized  \cite{Va12} that the composite system is represented by one state. Accordingly the system acts as a whole, giving rise to the momentum, that gets reflected in the AB quantum interference experiment. This non-localized interaction seems to be the explanation of this intriguing phenomenon from a classical physics perspective. The vector potential ${\mathbf A}$ may be still considered in this case as a convenient, intermediary mathematical step with $Q{\mathbf A}/c$ presenting a fa\c{c}ade of the mutual electric interaction of the conducting current carriers in the solenoid and an external charge $Q$, which gives rise to an  EM momentum in the system.
%---------------------------------------------
\section{Conclusions}
It was shown how in the AB double-slit interference experiment of electron beams, when a long but thin solenoid of electric current is introduced between the two slits, it gives rise to a subtle, equal and opposite EM momentum in the system associated with each electron beam. This causes an extra phase difference between the two interfering beams resulting in a shift in the interference pattern on the screen. The formal expression for the EM momentum attributes it to the presence of a vector potential, at the location of the charged beam, due to the solenoid of current even when there exists no magnetic field outside the solenoid. However, a viable physical explanation for the existence of this elusive momentum,  proportional to vector potential at the location of an external charged particle, has been unsuccessfully  coveted for long within classical electromagnetism. 
We showed that instead of looking at the effect of the vector potential at the external charge, if we consider the influence of the external charge on the sources of the vector potential, we could arrive at a satisfactory explanation of the elusive momentum.
%since inception of the idea more than 60 years back. 
Proceeding this way we showed how a small current loop, source of a finite vector potential at the location of an external electric charge, gives rise to a non-localized EM momentum. This momentum is manifested in the product of the drift velocities of the current-carrying charges in the current loop and the mass equivalent of their potential energies in the electric field of the external charge. We also showed that exactly the same EM momentum could also be computed alternatively if we instead employ the force due to electric field of the external electric charge on the current-carrying charges in the current loop. A pair of equal charges lying at symmetrical opposite locations outside the current loop gives rise to an equal and opposite EM momentum in the system. Supposedly it is this elusive, equal and opposite momentum for a pair of charges is what reflected through an extra phase difference between the interfering electron beams in the double-slit AB experiment, when a long but thin solenoid of current, comprising a stack of large number  of current loops, is introduced between the electron beams. It was further emphasized that one has to take the holistic view that the EM momentum actually lies in the composite system of the charges  plus solenoid, and that the composite system is represented by one state of the system. Accordingly the system acts as a whole, giving rise to the momentum, that gets reflected in the AB quantum interference experiment. This non-localized interaction seems to be the explanation of this intriguing phenomenon.
%A satisfactory explanation of this curious phenomenon in terms of such a subtle momentum under the aegis of classical electromagnetism, has remained elusive for long. Ostentatiously
%-------------------------------------------------------------------
\section*{Data Availability}
Data sharing not applicable to this article as no datasets were generated or analysed during the present study.
%--------------------------------------------
\section*{Author Declarations}
The author declares no conflicts/competing interests that are relevant to the content of this article.
\section*{Funding}
No funds or grants were received from anywhere for this research.
%%%%%%%%%%%%%%%%%%%%%%%%%%%%%%%%%%%%%%%%%%

%%%%%%%%%%%%%%%%%%%%%%%%%%%%%%%%%%%%%%%%%%%%%%%%%%%%%%%%%%%%%%%%%%%%%%%%%%%%%%%%%%
{}
\end{document}